\documentstyle[12pt,color,epsfig,subfigure]{article}
\def\baselinestretch{1.3}
\advance\textheight by \topskip
\newcommand{\comment}[1]{}

\def\beq{\begin{equation}}
\def\eeq{\end{equation}}
\def\beqn{\begin{eqnarray}}
\def\eeqn{\end{eqnarray}}

\begin{document}
\tolerance=100000
\thispagestyle{empty}
\setcounter{page}{0}
\topmargin -0.1in
\headsep 30pt
\footskip 40pt
\oddsidemargin 12pt
\evensidemargin -16pt
\textheight 8.5in
\textwidth 6.5in
\parindent 20pt
 
\def\baselinestretch{1.5}
\newcommand{\newc}{\newcommand}
\def\preprint{{preprint}}
\def\Ord{\lower .7ex\hbox{$\;\stackrel{\textstyle <}{\sim}\;$}}
\def\OOrd{\lower .7ex\hbox{$\;\stackrel{\textstyle >}{\sim}\;$}}
\def\cO#1{{\cal{O}}\left(#1\right)}
\newc{\order}{{\cal O}}
\def\lag             {{\cal L}}
\def\Lag             {{\cal L}}
\def\lum             {{\cal L}}
\def\R               {{\cal R}}
\def\Rsq             {{\cal R}^{\sq}}
\def\Rst             {{\cal R}^{\st}}
\def\Rsb             {{\cal R}^{\sb}}
\def\M               {{\cal M}}
\def\Oas             {{\cal O}(\alpha_{s})}
\def\Vcal            {{\cal V}}
\def\Wcal            {{\cal W}}
\newc{\be}{\begin{equation}}
\newc{\ee}{\end{equation}}
\newc{\br}{\begin{eqnarray}}
\newc{\er}{\end{eqnarray}}
\newc{\ba}{\begin{array}}
\newc{\ea}{\end{array}}
\newc{\bi}{\begin{itemize}}
\newc{\ei}{\end{itemize}}
\newc{\bn}{\begin{enumerate}}
\newc{\en}{\end{enumerate}}
\newc{\bc}{\begin{center}}
\newc{\ec}{\end{center}}
\newc{\ul}{\underline}
\newc{\ol}{\overline}
\newc{\ra}{\rightarrow}
\newc{\lra}{\longrightarrow}
\newc{\wt}{\widetilde}
\newc{\til}{\tilde}
\def\kr              {^{\dagger}}
\newc{\wh}{\widehat}
\newc{\ti}{\times}
\newc{\Dir}{\kern -6.4pt\Big{/}}
\newc{\Dirin}{\kern -10.4pt\Big{/}\kern 4.4pt}
\newc{\DDir}{\kern -10.6pt\Big{/}}
\newc{\DGir}{\kern -6.0pt\Big{/}}
\newc{\sig}{\sigma}
\newc{\sigmalstop}{\sig_{\lstoppair}}
\newc{\Sig}{\Sigma}  
\newc{\del}{\delta}
\newc{\Del}{\Delta}
\newc{\lam}{\lambda}
\newc{\Lam}{\Lambda}
\newc{\gam}{\gamma}
\newc{\Gam}{\Gamma}
\newc{\eps}{\epsilon}
\newc{\Eps}{\Epsilon}
\newc{\kap}{\kappa}
\newc{\Kap}{\Kappa}
\newc{\modulus}[1]{\left| #1 \right|}
\newc{\eq}[1]{(\ref{eq:#1})}
\newc{\eqs}[2]{(\ref{eq:#1},\ref{eq:#2})}
\newc{\etal}{{\it et al.}\ }
\newc{\ibid}{{\it ibid}.}
\newc{\ibidem}{{\it ibidem}.}
\newc{\eg}{{\it e.g.}\ }
\newc{\ie}{{\it i.e.}\ }
\def \viz{\emph{viz.}}
\def \etc{\emph{etc. }}
\newc{\nonum}{\nonumber}
\newc{\lab}[1]{\label{eq:#1}}
\newc{\dpr}[2]{({#1}\cdot{#2})}
\newc{\lt}{\stackrel{<}}
\newc{\gt}{\stackrel{>}}
\newc{\lsimeq}{\stackrel{<}{\sim}}
\newc{\gsimeq}{\stackrel{>}{\sim}}
\def\lsim{\buildrel{\scriptscriptstyle <}\over{\scriptscriptstyle\sim}}
\def\gsim{\buildrel{\scriptscriptstyle >}\over{\scriptscriptstyle\sim}}
\def\lapp{\mathrel{\rlap{\raise.5ex\hbox{$<$}}
                    {\lower.5ex\hbox{$\sim$}}}}
\def\gapp{\mathrel{\rlap{\raise.5ex\hbox{$>$}}
                    {\lower.5ex\hbox{$\sim$}}}}
\newc{\half}{\frac{1}{2}}
\newcommand {\nnc}        {{\overline{\mathrm N}_{95}}}
\newcommand {\dm}         {\Delta m}
\newcommand {\dM}         {\Delta M}
\def\bra{\langle}
\def\ket{\rangle}
\def\cO#1{{\cal{O}}\left(#1\right)}
\def \DM{{\Delta{m}}}
\newc{\bQ}{\ol{Q}}
\newc{\dota}{\dot{\alpha }}
\newc{\dotb}{\dot{\beta }}
\newc{\dotd}{\dot{\delta }}
\newc{\nindnt}{\noindent}

\newcommand{\medf}[2] {{\footnotesize{\frac{#1}{#2}} }}
\newcommand{\smaf}[2] {{\textstyle \frac{#1}{#2} }}
\def\onesq            {{\textstyle \frac{1}{\sqrt{2}} }}
\def\onehf            {{\textstyle \frac{1}{2} }}
\def\oneth            {{\textstyle \frac{1}{3} }}
\def\twoth            {{\textstyle \frac{2}{3} }}
\def\onefo            {{\textstyle \frac{1}{4} }}
\def\forth            {{\textstyle \frac{4}{3} }}

\newc{\matth}{\mathsurround=0pt}
\def\ML{\ifmmode{{\mathaccent"7E M}_L}
             \else{${\mathaccent"7E M}_L$}\fi}
\def\MR{\ifmmode{{\mathaccent"7E M}_R}
             \else{${\mathaccent"7E M}_R$}\fi}
\newcommand{\s}{\\ \vspace*{-3mm} }

\def \ud { {1 \over 2} }
\def \ut { {1 \over 3} }
\def \td { {3 \over 2} }
\newc{\mr}{\mathrm}
\def\dh {\partial }
\def \cs { cross-section }
\def \css { cross-sections }
\def \cm { centre of mass }
\def \cms { centre of mass energy }
\def \cc { coupling constant }
\def \ccs {coupling constants }
\def \gc {gauge coupling }
\def \gcc {gauge coupling constant }
\def \gccs {gauge coupling constants }
\def \yc {Yukawa coupling }
\def \ycc {Yukawa coupling constant }
\def \pp {{parameter }}
\def \pps {{parameters }} 
\def \ps {parameter space }
\def \pss {parameter spaces }
\def \vv {vice versa }

\newc{\siminf}{\mbox{$_{\sim}$ {\small {\hspace{-1.em}{$<$}}}    }}
\newc{\simsup}{\mbox{$_{\sim}$ {\small {\hspace{-1.em}{$>$}}}    }}


\newc {\Zboson}{{\mathrm Z}^{0}}
\newc{\thetaw}{\theta_W}
\newc{\mbot}{{m_b}}
\newc{\mtop}{{m_t}}
\newc{\sm}{${\cal {SM}}$}
\newc{\as}{\alpha_s}
\newc{\aem}{\alpha_{em}}
\def \PI{{\pi^{\pm}}}
\newc{\ppbar}{\mbox{$p\ol{p}$}}
\newc{\bbbar}{\mbox{$b\ol{b}$}}
\newc{\ccbar}{\mbox{$c\ol{c}$}}
\newc{\ttbar}{\mbox{$t\ol{t}$}}
\newc{\eebar}{\mbox{$e\ol{e}$}}
\newc{\zzero}{\mbox{$Z^0$}}
\def \gamz{\Gam_Z}
\newc{\wplus}{\mbox{$W^+$}}
\newc{\wminus}{\mbox{$W^-$}}
\newc{\ellp}{\ell^+}
\newc{\ellm}{\ell^-}
\newc{\elp}{\mbox{$e^+$}}
\newc{\elm}{\mbox{$e^-$}}
\newc{\elpm}{\mbox{$e^{\pm}$}}
\newc{\qbar}     {\mbox{$\ol{q}$}}
\def \ewgroup{SU(2)_L \otimes U(1)_Y}
\def \smgroup{SU(3)_C \otimes SU(2)_L \otimes U(1)_Y}
\def \smcolorem{SU(3)_C \otimes U(1)_{em}}

\def \SSM  {Supersymmetric Standard Model}
\def \poincare{Poincare$\acute{e}$}
\def \superspace{\emph{superspace}}
\def \sfs{\emph{superfields}}
\def \superpot{\emph{superpotential}}
\def \csf{\emph{chiral superfield}}
\def \csfs{\emph{chiral superfields}}
\def \vsf{\emph{vector superfield }}
\def \vsfs{\emph{vector superfields}}
\newc{\Ebar}{{\bar E}}
\newc{\Dbar}{{\bar D}}
\newc{\Ubar}{{\bar U}}
\newc{\susy}{{{SUSY}}}
\newc{\msusy}{{{M_{SUSY}}}}

\def\photino{\ifmmode{\mathaccent"7E \gam}\else{$\mathaccent"7E \gam$}\fi}
\def\taugluino{\ifmmode{\tau_{\mathaccent"7E g}}
             \else{$\tau_{\mathaccent"7E g}$}\fi}
\def\mphotino{\ifmmode{m_{\mathaccent"7E \gam}}
             \else{$m_{\mathaccent"7E \gam}$}\fi}
\newc{\gl}   {\mbox{$\wt{g}$}}
\newc{\mgl}  {\mbox{$m_{\gl}$}}
\def \charginopm{{\wt\chi}^{\pm}}
\def \mcharginopm{m_{\charginopm}}
\def \mchpmmin {\mcharginopm^{min}}
\def \chonep {{\wt\chi_1^+}}
\def \ch2p {{\wt\chi_2^+}}
\def \chonem {{\wt\chi_1^-}}
\def \ch2m {{\wt\chi_2^-}}
\def \chplus {{\wt\chi^+}}
\def \chminus {{\wt\chi^-}}
\def \chonip{{\wt\chi_i}^{+}}
\def \chonim{{\wt\chi_i}^{-}}
\def \chonipm{{\wt\chi_i}^{\pm}}
\def \chonjp{{\wt\chi_j}^{+}}
\def \chonjm{{\wt\chi_j}^{-}}
\def \chonjpm{{\wt\chi_j}^{\pm}}
\def \chonepm{{\wt\chi_1}^{\pm}}
\def \chonemp{{\wt\chi_1}^{\mp}}
\def \mchonepm{m_{\chonepm}}
\def \mchonemp{m_{\chonemp}}
\def \chtwopm{{\wt\chi_2}^{\pm}}
\def \mchtwopm{m_{\chtwopm}}
\newc{\dmchi}{\Delta m_{\wt\chi}}


\def \vlsp{\emph{VLSP}}
\def \lspi{\wt\chi_i^0}
\def \mlspi{m_{\lspi}}
\def \lspj{\wt\chi_j^0}
\def \mlspj{m_{\lspj}}
\def \lspone{\wt\chi_1^0}
\def \mlspone{m_{\lspone}}
\def \lsptwo{\wt\chi_2^0}
\def \mlsptwo{m_{\lsptwo}}
\def \lspthree{\wt\chi_3^0}
\def \mlspthree{m_{\lspthree}}
\def \lspfour{\wt\chi_4^0}
\def \mlspfour{m_{\lspfour}}


\newc{\sele}{\wt{\mathrm e}}
\newc{\sell}{\wt{\ell}}
\def \msell{m_{\sell}}
\def \slepone{\wt\ell_1}
\def \mslepone{m_{\slepone}}
\def \smuone{\wt\mu_1}
\def \msmuone{m_{\smuone}}
\def \stauone{\wt\tau_1}
\def \mstauone{m_{\stauone}}
\def \snu{\wt{\nu}}
\def \snutau{\wt{\nu}_{\tau}}
\def \msnu{m_{\snu}}
\def \msnumu{m_{\snu_{\mu}}}
\def \barsnu{\wt{\bar{\nu}}}
\def \barsnul{\barsnu_{\ell}}
\def \snul{\snu_{\ell}}
\def \mbarsnu{m_{\barsnu}}
\newc{\snue}     {\mbox{$ \wt{\nu_e}$}}
\newc{\smu}{\wt{\mu}}
\newc{\stau}{\wt{\tau}}
\newc {\nuL} {\wt{\nu}_L}
\newc {\nuR} {\wt{\nu}_R}
\newc {\snub} {\bar{\wt{\nu}}}
\newc {\eL} {\wt{e}_L}
\newc {\eR} {\wt{e}_R}
\def \slepl{\wt{l}_L}
\def \mslepl{m_{\slepl}}
\def \slepr{\wt{l}_R}
\def \mslepr{m_{\slepr}}
\def \stau{\wt\tau}
\def \mstau{m_{\stau}}
\def \slepton{\wt\ell}
\def \mslepton{m_{\slepton}}
\def \mlhiggs{m_{h^0}}

\def \xr{X_{r}}

\def \sfer{\wt{f}}
\def \msfer{m_{\sfer}}
\def \sq{\wt{q}}
\def \msq{m_{\sq}}
\def \msquleft{m_{\tilde{u_L}}}
\def \msqurht{m_{\tilde{u_R}}}
\def \sql{\wt{q}_L}
\def \msql{m_{\sql}}
\def \sqr{\wt{q}_R}
\def \msqr{m_{\sqr}}
\newc{\msqot}  {\mbox{$m_(\sq_{1,2} )$}}
\newc{\sqbar}    {\mbox{$\bar{\wt{q}}$}}
\newc{\ssb}      {\mbox{$\squark\ol{\squark}$}}
\newc {\qL} {\wt{q}_L}
\newc {\qR} {\wt{q}_R}
\newc {\uL} {\wt{u}_L}
\newc {\uR} {\wt{u}_R}
\def \ul{\wt{u}_L}
\def \mul{m_{\ul}}
\newc {\dL} {\wt{d}_L}
\newc {\dR} {\wt{d}_R}
\newc {\cL} {\wt{c}_L}
\newc {\cR} {\wt{c}_R}
\newc {\sL} {\wt{s}_L}
\newc {\sR} {\wt{s}_R}
\newc {\tL} {\wt{t}_L}
\newc {\tR} {\wt{t}_R}
\newc {\stb} {\ol{\wt{t}}_1}
\newc {\sbot} {\wt{b}_1}
\newc {\msbot} {m_{\sbot}}
\newc {\sbotb} {\ol{\wt{b}}_1}
\newc {\bL} {\wt{b}_L}
\newc {\bR} {\wt{b}_R}
\def \mul{m_{\wt{u}_L}}
\def \mur{m_{\wt{u}_R}}
\def \mdl{m_{\wt{d}_L}}
\def \mdr{m_{\wt{d}_R}}
\def \mcl{m_{\wt{c}_L}}
\def \charml{\wt{c}_L}
\def \mcr{m_{\wt{c}_R}}
\newc{\csquark}  {\mbox{$\wt{c}$}}
\newc{\csquarkl} {\mbox{$\wt{c}_L$}}
\newc{\mcsl}     {\mbox{$m(\csquarkl)$}}
\def \msl{m_{\wt{s}_L}}
\def \msr{m_{\wt{s}_R}}
\def \mbl{m_{\wt{b}_L}}
\def \mbr{m_{\wt{b}_R}}
\def \mtl{m_{\wt{t}_L}}
\def \mtr{m_{\wt{t}_R}}
\def \st{\wt{t}}
\def \mst{m_{\st}}
\newc {\stopl}         {\wt{\mathrm{t}}_{\mathrm L}}
\newc {\stopr}         {\wt{\mathrm{t}}_{\mathrm R}}
\newc {\stoppair}      {\wt{\mathrm{t}}_{1}
\bar{\wt{\mathrm{t}}}_{1}}
\def \lstop{\wt{t}_{1}}
\def \lstopbar{\lstop^*}
\def \hstop{\wt{t}_{2}}
\def \hstopbar{\hstop^*}
\def \mlstop{m_{\lstop}}
\def \mhstop{m_{\hstop}}
\def \lstoppair{\lstop\lstop^*}
\def \hstoppair{\hstop\hstop^*}
\newc{\tsquark}  {\mbox{$\wt{t}$}}
\newc{\ttb}      {\mbox{$\tsquark\ol{\tsquark}$}}
\newc{\ttbone}   {\mbox{$\tsquark_1\ol{\tsquark}_1$}}
\def \tsq {top squark }
\def \tsqs {top squarks }
\def \tsql {ligtest top squark }
\def \tsqh {heaviest top squark }
\newc{\mix}{\theta_{\wt t}}
\newc{\cost}{\cos{\theta_{\wt t}}}
\newc{\sint}{\sin{\theta_{\wt t}}}
\newc{\costloop}{\cos{\theta_{\wt t_{loop}}}}
\def \lsbot{\wt{b}_{1}}
\def \lsbotbar{\lsbot^*}
\def \hsbot{\wt{b}_{2}}
\def \hsbotbar{\hsbot^*}
\def \mlsbot{m_{\lsbot}}
\def \mhsbot{m_{\hsbot}}
\def \lsbotpair{\lsbot\lsbot^*}
\def \hsbotpair{\hsbot\hsbot^*}
\newc{\mixsbot}{\theta_{\wt b}}

\def \mhone{m_{h_1}}
\def \hup{{H_u}}
\def \hdn{{H_d}}
\newc{\tb}{\tan\beta}
\newc{\cb}{\cot\beta}
\newc{\vev}[1]{{\left\langle #1\right\rangle}}

\def \abot{A_{b}}
\def \atop{A_{t}}
\def \atau{A_{\tau}}
\newc{\mhalf}{m_{1/2}}
\newc{\mzero} {\mbox{$m_0$}}
\newc{\azero} {\mbox{$A_0$}}

\newc{\lb}{\lam}
\newc{\lbp}{\lam^{\prime}}
\newc{\lbpp}{\lam^{\prime\prime}}
\newc{\rpv}{{\not \!\! R_p}}
\newc{\rpvm}{{\not  R_p}}
\newc{\rp}{R_{p}}
\newc{\rpmssm}{{RPC MSSM}}
\newc{\rpvmssm}{{RPV MSSM}}


\newc{\sbyb}{S/$\sqrt B$}
\newc{\pelp}{\mbox{$e^+$}}
\newc{\pelm}{\mbox{$e^-$}}
\newc{\pelpm}{\mbox{$e^{\pm}$}}
\newc{\epem}{\mbox{$e^+e^-$}}
\newc{\lplm}{\mbox{$\ell^+\ell^-$}}
\def \branch{\emph{BR}}
\def \branche{\branch(\lstop\ra be^{+}\nu_e \lspone)\ti \branch(\lstop^{*}\ra \bar{b}q\bar{q^{\prime}}\lspone)}
\def \branchmu{\branch(\lstop\ra b\mu^{+}\nu_{\mu} \lspone)\ti \branch(\lstop^{*}\ra \bar{b}q\bar{q^{\prime}}\lspone)}
\def\Ecm{\ifmmode{E_{\mathrm{cm}}}\else{$E_{\mathrm{cm}}$}\fi}
\newc{\rts}{\sqrt{s}}
\newc{\rtshat}{\sqrt{\hat s}}
\newc{\gev}{\,GeV}
\newc{\mev}{~{\rm MeV}}
\newc{\tev}  {\mbox{$\;{\rm TeV}$}}
\newc{\gevc} {\mbox{$\;{\rm GeV}/c$}}
\newc{\gevcc}{\mbox{$\;{\rm GeV}/c^2$}}
\newc{\intlum}{\mbox{${ \int {\cal L} \; dt}$}}
\newc{\call}{{\cal L}}
\def \met  {\mbox{${E\!\!\!\!/_T}$}}
\def \cpv  {\mbox{${CP\!\!\!\!/}$}}
\newc{\ptmiss}{/ \hskip-7pt p_T}
\def \eslash{\not \! E}
\def \etslash{\not \! E_T }
\def \ptslash{\not \! p_T }
\newc{\PT}{\mbox{$p_T$}}
\newc{\ET}{\mbox{$E_T$}}
\newc{\dedx}{\mbox{${\rm d}E/{\rm d}x$}}
\newc{\ifb}{\mbox{${\rm fb}^{-1}$}}
\newc{\ipb}{\mbox{${\rm pb}^{-1}$}}
\newc{\pb}{~{\rm pb}}
\newc{\fb}{~{\rm fb}}
\newc{\ycut}{y_{\mathrm{cut}}}
\newc{\chis}{\mbox{$\chi^{2}$}}
\def \hadron{\emph{hadron}}
\def \nlc{\emph{NLC }}
\def \lhc{\emph{LHC }}
\def \cdf{\emph{CDF }}
\def\dzero{\emptyset}
\def \tevatron{\emph{Tevatron }}
\def \lep{\emph{LEP }}
\def \jets{\emph{jets }}
\def \jet(s){\emph{jet(s) }}

\def\Crs{stroke [] 0 setdash exch hpt sub exch vpt add hpt2 vpt2 neg V currentpoint stroke 
hpt2 neg 0 R hpt2 vpt2 V stroke}
\def\loopdk{\lstop \ra c \lspone}
\def\brloopdk{\branch(\loopdk)}
\def\fourdk{\lstop \ra b \lspone  f \bar f'}
\def\brfourdk{\branch(\fourdk)}
\def\fourdklep{\lstop \ra b \lspone  \ell \nu_{\ell}}
\def\fourdkhad{\lstop \ra b \lspone  q \bar q'}
\def\brfourdklep{\branch(\fourdklep)}
\def\brfourdkhad{\branch(\fourdkhad)}
\def\twodk{\lstop \ra b \chonep}
\def\brtwodk{\branch(\twodk)}
\def\threedkslep{\lstop \ra b \wt{\ell} \nu_{\ell}}
\def\brthreedkslep{\branch(\threedkslep)}
\def\threedksnu{\lstop \ra b \wt{\nu_{\ell}} \ell }
\def\brthreedksnu{\branch(\threedksnu) }
\def\threedklsp{\lstop \ra b W \lspone }
\def\brthreedklsp{\\branch(\threedklsp) }
\def\topdk{t \ra \lstop \lspone}
\def\rpvdk{\lstop \ra e^+ d}
\def\brrpvdk{\branch(\rpvdk)}
\def\fonec{f_{11c}} 
\newc{\mpl}{M_{\rm Pl}}
\newc{\mgut}{M_{GUT}}
\newc{\mw}{M_{W}}
\newc{\mweak}{M_{weak}}
\newc{\mz}{M_{Z}}

\newc{\OPALColl}   {OPAL Collaboration}
\newc{\ALEPHColl}  {ALEPH Collaboration}
\newc{\DELPHIColl} {DELPHI Collaboration}
\newc{\XLColl}     {L3 Collaboration}
\newc{\JADEColl}   {JADE Collaboration}
\newc{\CDFColl}    {CDF Collaboration}
\newc{\DXColl}     {D0 Collaboration}
\newc{\HXColl}     {H1 Collaboration}
\newc{\ZEUSColl}   {ZEUS Collaboration}
\newc{\LEPColl}    {LEP Collaboration}
\newc{\ATLASColl}  {ATLAS Collaboration}
\newc{\CMSColl}    {CMS Collaboration}
\newc{\UAColl}    {UA Collaboration}
\newc{\KAMLANDColl}{KamLAND Collaboration}
\newc{\IMBColl}    {IMB Collaboration}
\newc{\KAMIOColl}  {Kamiokande Collaboration}
\newc{\SKAMIOColl} {Super-Kamiokande Collaboration}
\newc{\SUDANTColl} {Soudan-2 Collaboration}
\newc{\MACROColl}  {MACRO Collaboration}
\newc{\GALLEXColl} {GALLEX Collaboration}
\newc{\GNOColl}    {GNO Collaboration}
\newc{\SAGEColl}  {SAGE Collaboration}
\newc{\SNOColl}  {SNO Collaboration}
\newc{\CHOOZColl}  {CHOOZ Collaboration}
\newc{\PDGColl}  {Particle Data Group Collaboration}

\def\issue(#1,#2,#3){{\bf #1}, #2 (#3)}
\def\ASTR(#1,#2,#3){Astropart.\ Phys. \issue(#1,#2,#3)}
\def\AJ(#1,#2,#3){Astrophysical.\ Jour. \issue(#1,#2,#3)}
\def\AJS(#1,#2,#3){Astrophys.\ J.\ Suppl. \issue(#1,#2,#3)}
\def\APP(#1,#2,#3){Acta.\ Phys.\ Pol. \issue(#1,#2,#3)}
\def\JCAP(#1,#2,#3){Journal\ XX. \issue(#1,#2,#3)} 
\def\SC(#1,#2,#3){Science \issue(#1,#2,#3)}
\def\PRD(#1,#2,#3){Phys.\ Rev.\ D \issue(#1,#2,#3)}
\def\PR(#1,#2,#3){Phys.\ Rev.\ \issue(#1,#2,#3)} 
\def\PRC(#1,#2,#3){Phys.\ Rev.\ C \issue(#1,#2,#3)}
\def\NPB(#1,#2,#3){Nucl.\ Phys.\ B \issue(#1,#2,#3)}
\def\NPPS(#1,#2,#3){Nucl.\ Phys.\ Proc. \ Suppl \issue(#1,#2,#3)}
\def\NJP(#1,#2,#3){New.\ J.\ Phys. \issue(#1,#2,#3)}
\def\JP(#1,#2,#3){J.\ Phys.\issue(#1,#2,#3)}
\def\PL(#1,#2,#3){Phys.\ Lett. \issue(#1,#2,#3)}
\def\PLB(#1,#2,#3){Phys.\ Lett.\ B  \issue(#1,#2,#3)}
\def\ZP(#1,#2,#3){Z.\ Phys. \issue(#1,#2,#3)}
\def\ZPC(#1,#2,#3){Z.\ Phys.\ C  \issue(#1,#2,#3)}
\def\PREP(#1,#2,#3){Phys.\ Rep. \issue(#1,#2,#3)}
\def\PRL(#1,#2,#3){Phys.\ Rev.\ Lett. \issue(#1,#2,#3)}
\def\MPL(#1,#2,#3){Mod.\ Phys.\ Lett. \issue(#1,#2,#3)}
\def\RMP(#1,#2,#3){Rev.\ Mod.\ Phys. \issue(#1,#2,#3)}
\def\SJNP(#1,#2,#3){Sov.\ J.\ Nucl.\ Phys. \issue(#1,#2,#3)}
\def\CPC(#1,#2,#3){Comp.\ Phys.\ Comm. \issue(#1,#2,#3)}
\def\IJMPA(#1,#2,#3){Int.\ J.\ Mod. \ Phys.\ A \issue(#1,#2,#3)}
\def\MPLA(#1,#2,#3){Mod.\ Phys.\ Lett.\ A \issue(#1,#2,#3)}
\def\PTP(#1,#2,#3){Prog.\ Theor.\ Phys. \issue(#1,#2,#3)}
\def\RMP(#1,#2,#3){Rev.\ Mod.\ Phys. \issue(#1,#2,#3)}
\def\NIMA(#1,#2,#3){Nucl.\ Instrum.\ Methods \ A \issue(#1,#2,#3)}
\def\JHEP(#1,#2,#3){J.\ High\ Energy\ Phys. \issue(#1,#2,#3)}
\def\EPJC(#1,#2,#3){Eur.\ Phys.\ J.\ C \issue(#1,#2,#3)}
\def\RPP (#1,#2,#3){Rept.\ Prog.\ Phys. \issue(#1,#2,#3)}
\def\PPNP(#1,#2,#3){ Prog.\ Part.\ Nucl.\ Phys. \issue(#1,#2,#3)}
\newc{\PRDR}[3]{{Phys. Rev. D} {\bf #1}, Rapid  Communications, #2 (#3)}

\vspace*{\fill}
\vspace{-1.2in}
\begin{flushright}
{\tt IISER/HEP/07/08}
\end{flushright}
\begin{center}
{\Large \bf
Lepton Flavours at the Early LHC Experiments as the Footprints
of the Dark Matter Producing Mechanisms
}
  \vglue 0.5cm
Nabanita Bhattacharyya$^{(a)}$\footnote{nabanita@iiserkol.ac.in},
Amitava Datta$^{(a)}$\footnote{adatta@iiserkol.ac.in} and
Sujoy Poddar$^{(a)}$\footnote{sujoy$\_$phy@iiserkol.ac.in}
    \vglue 0.2cm
    {\it $^{(a)}$
Indian Institute of Science Education and Research, Kolkata, \\
HC-VII,Sector III, Salt Lake City, Kolkata 700 106, India.
\\}
\end{center}
\vspace{.2cm}


\begin{abstract}
{\noindent \normalsize}

The mSUGRA parameter space corresponding to light sleptons well within 
the reach of LHC and relatively 
light squarks and gluinos (mass $\le$ 1 TeV) has three regions consistent 
with the WMAP data on dark matter relic density and direct mass bounds 
from LEP 2. Each region can lead to distinct leptonic signatures from 
squark-gluino events during the early LHC experiments 
(integrated luminosity $\sim 10 ~\ifb$ or even smaller). In the much studied 
stau-LSP coannihilation region with a vanishing 
common trilinear coupling ($A_0$) at the GUT scale a large fraction of the final states contain electrons 
and / or muons and $e$ - $\mu$ - $\tau$ universality holds to a good approximation. In the 
not so well studied scenarios with non-vanishing $A_0$ both LSP pair 
annihilation and stau-LSP coannihilation could contribute significantly to 
the dark matter relic density for even smaller squark-gluino masses. Our 
simulations indicate that the corresponding signatures are final states 
rich in $\tau$-leptons while final states with electrons and 
muons are suppressed leading to a violation of lepton universality.
These features may be observed to a lesser extent even in the modified parameter 
space (with non-zero $A_0$) where the coannihilation process dominates. We also show 
that the generic $m$-leptons + $n$-jets+ $\met$ signatures without 
flavour tagging can also discriminate among the three scenarios. However, 
the signals become more informative if the $\tau$ and $b$-jet tagging 
facilities at the LHC experiments are utilized.  
\end{abstract}

\newpage

\section{Introduction}\label{intro4} 
~~~Models with supersymmetry(SUSY) \cite{SUSY} are interesting for a variety of
theoretical and phenomenological reasons.  
A specially attractive
feature of the minimal supersymmetric standard model (MSSM) with R-parity
conservation is the presence of the stable, weakly interacting lightest
neutralino (${\tilde \chi}_1^0$ ) \cite{goldbergDM} which is assumed to 
the lightest supersymmetric particle (LSP). This turns  out to be a
very good candidate for the observed dark matter (DM) in the universe
\cite{DMreviewgeneral,recentSUSYDMreview,DMreview}.

Various SUSY models have been proposed and constrained by the data on DM
relic density \cite{coannistau}-\cite{funnel}. The recent revival of interest in this field is due 
to the very restrictive data from the Wilkinson Microwave Anisotropy Probe 
(WMAP) observation \cite{WMAPdata}. Combining the WMAP data with the results 
from the SDSS (Sloan Digital Sky Survey) one obtains the conservative 3 
$\sigma$ limits
\begin{equation}
0.09 < \Omega_{DM}h^2 < 0.13
\label{relicdensity}
\end{equation}

\noindent 
where $\Omega_{DM}h^2$ is is the DM relic density in units of 
the critical density, $h=0.71\pm0.026$ is the Hubble constant in units of 
$100 \ \rm Km \ \rm s^{-1}\ \rm Mpc^{-1}$. In this paper
we shall assume that $\Omega_{DM}\equiv \Omega_{{\tilde \chi}_1^0}$. We 
should note here that 
the upper bound on $\Omega_{{\tilde \chi}_1^0}$ in Eq.(\ref{relicdensity}) 
must hold in any model with SUSY. In contrast the 
lower bound evaporates if the possibility of
non-SUSY origin of DM is left open.

In the thermally generated DM scenario the present value of 
$\Omega_{{\tilde \chi}_1^0}h^2$ can be computed by solving the Boltzmann 
equation for $n_{{\tilde \chi}_1^0}$, the number density of the LSP in a 
Friedmann-Robertson-Walker universe. The most important particle physics 
input in this calculation is the thermally averaged quantity 
$<\sigma_{eff} v>$, where $v$ is the relative velocity between two 
neutralinos annihilating each other and $\sigma_{eff}$ is the annihilation 
cross-section for all possible final states involving SM particles only. 
In addition to the negation of a LSP pair, coannihilation of the LSP 
\cite{coannistau}-\cite{coanniSet2},\cite{coannistop} with
supersymmetric particles 
(sparticles) approximately degenerate with the LSP may also be important. 
The smaller the annihilation/coannihilation cross-section the larger 
becomes the LSP relic density.
In addition to the parameters of the standard model the 
annihilation cross-section $\sigma_{eff}$ depends on the masses and the 
couplings of the sparticles and on the magnitudes of the 
bino ($\tilde B$), wino ($\tilde W$) and Higgsino (${\tilde H}_1^0$ 
, ${\tilde H}_2^0$) components of the LSP. Discovery of SUSY at the LHC 
followed by the measurement of the above parameters can, therefore, verify 
the hypothesis of supersymmetric DM as well as identify the underlying DM 
relic density producing mechanism \cite{dmlhc}. This, however, is likely to take some time.

SUSY models with relatively small squark, gluino masses are of 
considerable contemporary interest since such strongly interacting 
sparticles with large production cross-sections are expected to show up in 
the early stages of the LHC experiments. In this case there are a few 
important relic density producing mechanisms like the LSP pair 
annihilation, LSP - lighter stau ($\stauone$) and LSP-lighter-stop 
($\lstop$) coannihilation \cite{dreesDM93}, 
\cite{coannistau}-\cite{coanniSet2}, \cite{coannistop}. Each mechanism is 
active in one or more regions of the SUSY parameter space. It is 
worthwhile to check whether the sparticle spectra corresponding to each 
region yield distinct signatures at the early LHC experiments so that some 
idea of the DM producing mechanism, though qualitative, can be obtained.

In this paper we shall restrict ourselves to the popular minimal 
supergravity (mSUGRA)model \cite{msugra} with  moderate values of the 
parameter tan $\beta$ (to be defined in the next section). Our attention 
will be focussed on the regions of the parameter space consistent with the 
WMAP data and within the reach of the early runs. This region necessarily 
corresponds to light sleptons well within the reach of LHC 
but unlikely to be discovered directly by the early experiments (see section 3 for 
the details). Moreover the slepton signature alone carries very little 
information about the underlying DM relic density producing mechanism.

In mSUGRA the above scenario also involves relatively light squarks and 
gluinos. The lighter chargino ($\chonepm$) and the second lightest 
neutralino ($\lsptwo$) are copiously present in the squark-gluino decay 
cascades. In the light slepton scenario these inos almost exclusively 
decay leptonically via two body decay modes. The lepton flavour content of 
the final states thus obtained from squark-gluino events at the early 
stages of the LHC run may reflect the underlying relic density producing 
mechanisms \cite{debottam} (see section 2 for a brief review). Thus at least 
the footprints of these mechanisms may be viewed long before  
reconstruction of the sparticle masses and other relevant parameters 
establish the model rigorously.

In the next section we shall review the important DM relic density 
producing mechanisms in different regions of the mSUGRA parameter space
with sparticle spectrum as described above and qualitatively review the 
characteristics of the squark-gluino signatures from each region.
In section 3 we shall go beyond \cite{debottam} and present the results 
of our simulations revealing new aspects of the signals. This will
justify the qualitative discussions of the previous section. 
The summary along with future outlooks will be  the content of the last section.

\section{The Early LHC signatures: a qualitative discussion}

~~~The simplest gravity mediated SUSY breaking model - the minimal 
supergravity (mSUGRA) \cite{msugra} model- has only five free parameters. 
These are $m_{1/2}$ ( the common gaugino mass), $m_0$ (the common scalar mass) and 
the common trilinear coupling parameter $A_0$, all given at the 
gauge coupling unification scale ($M_G \sim 2 \times 10^{16}$~GeV), the 
ratio of Higgs vacuum expectation values at the electroweak scale namely 
$\tan\beta$ and the sign of 
$\mu$, the higgsino mixing parameter. The magnitude of $\mu$ is determined by 
the radiative electroweak symmetry breaking (REWSB) condition
\cite{rewsbrefs}. The low energy sparticle spectra and couplings 
at the electroweak scale are generated by renormalization group evolutions 
(RGE) of the soft breaking masses and coupling parameters \cite{oldrge}. 
In this paper we shall restrict ourselves to a moderate value 
of tan $\beta$ namely 10 and positive 
$\mu $. Representative values of the remaining parameters will be used to highlight different
DM relic density producing mechanisms and the corresponding collider signals.
Since the entire sparticle spectra and the couplings can be computed in 
terms of the five parameters only,
the calculation of the LSP annihilation or coannihilation cross-sections 
and, consequently, the DM relic density are rather precise in this 
framework \cite{coannistau,dreesDM93,dmsugra}.


The WMAP allowed regions of the mSUGRA parameter space can be
classified into several regions depending on the dominant  
LSP annihilation/coannihilation mechanisms. The details can be found in 
\cite{DMreviewgeneral,recentSUSYDMreview,DMreview,dmsugra}.
In the following
we list the mechanisms which will be relevant for the discussions in 
this paper focussing on relatively light sleptons, squarks and gluinos.

One such region corresponds to small $m_0$ but somewhat larger $\mhalf$ 
(numerical examples will be given later). This choice leads to sleptons 
lighter than the charginos and the heavier neutralinos. Moreover, the mass 
difference between the lighter stau ($\stauone$) and the LSP turns out to 
be at most 30 GeV or so. Consequently 
$\stauone$ -LSP coannihilation \cite{coannistau} often abbreviated as  
$\stau$-coannihilation occurs quite efficiently sufficient for
producing the DM in the universe. The importance of this 
region, which we shall refer to as the conventional $\stau$- 
coannihilation zone, in the context of the WMAP data has recently 
been emphasized by several groups \cite{dmsugra}.

Most of the above analyses, however, were based on the choice $A_0 = $ 0
without any compelling theoretical
or  empirical reason. One  consequence of this ad hoc choice is that in 
order to satisfy the LEP bound on the lightest Higgs boson mass: $m_h> 114.4$ GeV \cite{limits}  
one requires relatively large $\mhalf$. Typically for $m_{0} \sim 100$ GeV one requires $m_{1/2} \sim 500$ GeV
to satisfy the $m_h$ bound and the WMAP constraints. As a result the squarks 
and gluinos become approximately degenerate, the latter  being slightly 
heavier. The masses of these superpartners turn out to be $\order$(1 TeV) 
or larger. Nevertheless the total squark-gluino production cross 
section is sufficient to produce observable signatures at the LHC
(See, e.g., \cite{tata}) 
throughout the $\stau$-coannihilation region.
The lighter top squark mass is smaller than the other squarks due to the 
usual renormalization group 
effects driven by its large Yukawa coupling. However, due to the 
choice $A_0$ = 0 its mass is not further suppressed by mixing 
effects in the mass matrix. Therefore its mass is also of the order of
one TeV. Such heavy $\lstop$ does not participate in $\lstop$ - LSP coannihilation.
Moreover the production of $\lstop$ pairs do not affect the total 
squark-gluino production significantly.
The features discussed in the last two paragraphs lead to distinct collider 
signatures as we shall illustrate  below with numerical examples.

The other region of interest is the  bulk annihilation region or the 
bulk region \cite{recentSUSYDMreview,DMreview,dreesDM93} where 
$m_0$ and $m_{1/2}$ are such that many of the sparticles are significantly 
lighter than those in the conventional $\stau$-coannihilation region. The 
LSP turns 
out to be bino dominated and, consequently, couples favourably to right 
sleptons, 
which in fact are the lightest sfermions in this region of the parameter 
space. As a result an LSP pair efficiently annihilates into SM fermions 
via the exchange of light sfermions in the t-channel. This cross-section 
depends on the mass of the LSP ($\mlspone$), the masses of the exchanged sfermions
and the LSP-sfermion couplings \cite{recentSUSYDMreview,DMreview,dreesDM93}. This region characterized by 
relatively light sparticles is especially interesting since SUSY signals 
with large events rates may be expected at the early LHC runs.

Strong lower bounds on sparticle masses \cite{hlim}, particularly on the 
slepton masses from LEP 2 disfavor a part of the bulk annihilation zone. 
We, however, wish to emphasize that the direct bounds on the slepton 
masses alone can not eliminate the entire bulk region. Nevertheless a more 
severe restriction practically rules out the $(m_0-m_{1/2})$ plane 
containing the bulk region for $A_0 =$ 0. This arises from the 
LEP 2 bound on lightest 
Higgs boson mass ($m_h$) \cite{limits} since 
$m_h$ and slepton masses are correlated in mSUGRA. 
Thus it has often been claimed in the 
recent literature that the mSUGRA parameter space with low values of both 
$m_0$ and $\mhalf$ and, consequently, the entire bulk region is strongly 
disfavoured \cite{dmsugra}.

It was emphasized in \cite{debottam} that the above conclusions are 
artifacts of the ad hoc choice $A_0 = 0$.  
On the other hand it is well known that for given $ m_0$ and $m_{1/2}$ moderate to 
large negative\footnote{ We follow the standard sign convention of 
Ref. \cite{signconvention} for the signs of $\mu$ and $A_0$.} values of 
$A_0$ lead to larger $m_h $ $ \cite{carenaEtc}$\footnote{For $A_0>0$, one requires $m_0$
and $\mhalf$ typically larger than the corresponding values for $A_0$=0.
This does not lead to any novel collider signal.}. 
Hence in this case the bound 
on $m_h$ can be satisfied even for relatively small $m_0$ and $\mhalf$. 
This revives the region where LSP pair annihilation is the 
dominant DM producing mechanism.
This can be seen, e.g.,from Fig.1 of \cite{debottam} 
(see the blue(deep shaded)region; several other figures of the above 
reference corresponding to different choices of mSUGRA parameters also 
exhibit similar features). Moreover the low 
$m_0-m_{1/2}$ regions of the mSUGRA parameter space are characterized by 
relatively light squarks and gluinos. This along with the inevitable presence of 
light sleptons in this region of the parameter space leads to distinct
signals from squark-gluino events.

There is also a region where $\stau$-coannihilation is still the 
most important mechanism for creating the observed DM in the 
universe( see the pink (the light shaded) regions of Fig. 1 \cite{debottam}
and other similar
figures). Remarkably, even this region corresponds to  much smaller 
$\mhalf$ compared to what one would obtain for the $A_0$ =0 case (i.e., 
for the conventional $\stau$-coannihilation region).

Furthermore large negative values of $A_0$ leads to a relatively light top 
squark. In fact for a small but non-negligible region of the parameter 
space, the LSP - $\lstop$ coannihilation \cite{coannistop} along with bulk 
annihilation may significantly contribute to the observed DM
density with the $\lstop$ well within the kinematic reach of the Tevatron 
(see the red region of Fig. 1 \cite{debottam} and other figures).

In \cite{debottam} some features of the sparticle spectrum and signals at 
the Tevatron and the LHC corresponding to the WMAP allowed regions of the 
parameter space opened up by non-zero $A_0$ were studied. The results were 
compared and contrasted with the expectations from the well publicized 
conventional $\stauone$-coanihilation scenario with $A_0 = 0$
by introducing three benchmark scenarios A, B and C ( 
Table 1 of \cite{debottam}, which has also been reproduced here as Table 1 
for a ready reference). In scenario A with a relatively large $A_0$ both 
bulk annihilation and  $\stau$-coannihilation are responsible for 
producing the observed DM relic density although the former 
dominates. On the other hand in scenario B with somewhat larger $\mhalf$
the latter dominates but the 
contribution of the bulk annihilation is non-negligible. Finally in 
scenario C with $A_0 =$ 0 the conventional $\stau$-
coannihilation is the only 
significant DM producing mechanism.
The scenario C corresponds to the 
smallest $\mhalf$ 
which is consistent with the Higgs mass bound and the WMAP data. 

In the following the main features  of the signal in the three scenarios 
\cite{debottam} are summarized. The tables referred to in the rest of this 
section belongs to the original work unless stated otherwise explicitly.

The sparticle spectra in the three scenarios can be found in Table 2 and 
the total squark-gluino cross-section in Table 6.
These lowest order cross-sections have been computed
by CalcHEP (version 2.3.7) \cite{calchep}.

The signals at the LHC are governed by the cascade decays of the above 
sparticles. In all three cases the gluinos are heavier than all squarks. 
As a result the gluinos decay into quark-squark pairs (see Table 3) . The 
squarks belonging to the third generation are relatively light due to the 
usual renormalization group effects in the mSUGRA model. Their masses are 
further suppressed by mixing effects in the mass matrix due to the 
non-vanishing $A_0$ parameter scenarios A and B. As a result 
these squarks are more frequently present in gluino decay products in 
these scenarios compared to C. 

The squarks belonging to the first two generations in general decay into 
the corresponding lighter quarks and an appropriate electroweak gaugino 
(Table 3). The lighter top squark similarly decays into appropriate 
quark-gaugino pairs. The bottom squarks of both type may decay, in 
addition to above channels, into a lighter top squark and a W boson in 
some scenarios (Table 4) with non-negligible branching ratios(BRs). 
Nevertheless, the decay of each third generation squark inevitably 
contains a b quark. This is the origin of the large fraction of final 
states with b-partons as noted in \cite{debottam}. In scenario C the 
fraction of third generation squarks in gluino decay is relatively small 
and the above effect is suppressed to some extent.

The decay properties of the lighter chargino ($ \chonepm$)and the second lightest 
neutralino ($\lsptwo $) which, in addition to the LSP, are often present in squark-gluino 
decays determine the lepton content of the final sates to 
a large extent. As a  direct consequence of the presence of the light 
sleptons,  these two 
unstable gauginos decay almost exclusively into leptonic channels via two body 
modes in all three scenarios. 

In Table 2 of this paper, which is an enlarged version of 
Table 5 of \cite{debottam}, we present the relevant branching 
ratios of $ \chonepm$ and $\lsptwo $. In scenario A the wino dominated lighter chargino, is not 
kinematically allowed to decay into L-type charged sleptons. Its two body 
decay into a lepton-sneutrino pair, though kinematically allowed, is phase 
space suppressed. It therefore decays into R-type sleptons through its 
subdominant Higgsino component with a large BR. This results in a very 
large fraction of final states containing the $\stauone$, which 
eventually 
decays into a $\tau$-LSP  pair. Only a tiny fraction of the 
final states contains electrons and muons. Unlike the SM the hallmark of 
this scenario is, 
therefore, lepton non-universality in the final states. 
For similar reasons the $\lsptwo$ decays primarily into $\tau$-$\stauone$ 
pair while a much smaller fraction decays into neutrino-sneutrino pairs, 
contributing further to the $\tau$ dominance of the final states. The 
sneutrinos in turn decay into the invisible neutrino-LSP channel in all 
three scenarios and act as additional carriers of missing energy.

The scenario B has all the above features albeit to a lesser extent. The $\tau$ 
dominance in both $\chonepm$ and $\lsptwo$ decays exists but is reduced 
significantly compared to the predictions of scenario A. The fraction of 
invisible decays of $ \lsptwo$ into $ \nu - \snu$ pair increases. This common 
feature of A and B can be easily illustrated by a parton level calculation 
(see Table 7). In a realistic LHC experiment with good $\tau$ tagging 
capabilities the observability of this $\tau$ excess has been demonstrated by 
simulation(see Table 8) using Pythia.

In Table 2 our focus was on the three bench mark scenarios. In order to
illustrate  that the lepton flavour content of the final state is indeed 
correlated with the DM producing mechanism we have selected several 
representative points ($S_{1} - S_{7}$) from the figures in \cite{debottam}. 
The dominant DM producing mechanism and the figure no. in \cite{debottam} are given in parentheses.
For all points tan$\beta = 10 ; \mu > 0  $. The other parameters are as given below.\\

\noindent
$S_{1}$ : $m_0 = 100 ; m_{1/2} = 250 ; A_{0} = -700 $ (Bulk ; Fig. 4)\\                                                                                                                              
$S_{2}$ : $m_0 = 125 ; m_{1/2} = 400 ; A_{0} = -700  $ (Coannihilation ; Fig. 4)\\                                                                                                                              
$S_{3}$ : $m_0 = 120 ; m_{1/2} = 400 ; A_{0} = -800  $ (Coannihilation ; Fig. 1(a))\\                                                                                                                              
$S_{4}$ : $m_0 = 120 ; m_{1/2} = 265 ; A_{0} = -900  $ (Bulk ; Fig. 1(a))\\                                                                                                                              
$S_{5}$ : $m_0 = 120 ; m_{1/2} = 300 ; A_{0} = -1000 $ (Bulk ; Fig. 1(a))\\                                                                                                                              
$S_{6}$ : $m_0 = 170 ; m_{1/2} = 500 ; A_{0} = -1200 $ (Coannihilation ; Fig. 1(c))\\                                                                                                                              
$S_{7}$ : $m_0 = 170 ; m_{1/2} = 500 ; A_{0} = -1400 $ (Coannihilation ; Fig. 1(c)) \\ 

The relevant BRs of lighter chargino and second lightest neutralino decays
are presented in Table 3. It is readily seen that if the bulk
annihilation ($\stau$-coannihilation ) is the dominant mechanism, 
the decays follow the patterns of scenario A (B) (compare with Table 2).


In scenario C the $\chonepm$ decays into left slepton- neutrino pairs 
or sneutrino -lepton pairs belonging to the first two generations with 
equal BR of sizable magnitudes. The fraction of final states 
involving $\tau$s is only marginally larger and lepton universality holds 
to a very good approximation. In fact due to limited $\tau$ detection 
efficiency, the fraction of observed final states involving $e$ and/or $\mu$ 
will be apparently larger in stark contrast to the predictions of scenarios A and B. 
The decays of $\lsptwo$ contributes further to the restoration of lepton 
universality. The $\lsptwo$ now dominantly decays into the invisible final state 
consisting of a neutrino and a sneutrino with 51.6 \% BR. Thus it largely 
acts as a carrier of missing energy in addition to the LSP and the 
sneutrino. Although we have presented numerical results for scenario C only
we have verified that the above features hold qualitatively for the entire $\stau$-coannihilation strip. 

Scenarios with the electroweak gauginos decaying into purely leptonic two 
body channels and sneutrinos and $\lsptwo$ acting as additional carriers 
of missing energy have already been discussed in details in the context of 
MSSM. The characteristic signatures from final states containing excess of 
electrons and /or muons and more than one invisible particles at the 
Tevatron \cite{admgbm} and at $e^+ - e^-$ colliders like LEP or the NLC \cite{addrees,adaksr} were 
simulated at the parton level. An ISAJET based analysis in the context of 
the Tevatron Run I was also done \cite{adparua}. The scope of accommodating 
such scenarios in the mSUGRA models and several of its variants was also 
discussed \cite{adparida}.

However, the connection of this scenario with a  WMAP allowed region of 
the mSUGRA parameter space has not been highlighted in the existing 
literature. Nor were the LHC signatures of this scenario studied with due 
emphasis. It is interesting to note that this virtual LSP(`VLSP') or 
effective LSP(`ELSP') scenario\footnote{ The invisibly decaying $\wt 
{\nu}$'s and $\lsptwo$ acting as additional carriers of missing energy
may be  called VLSP or ELSP in the context of collider 
physics} is 
realized in the well studied conventional $\stau$-coannihilation region of 
the mSUGRA parameter space. In the next section we shall simulate the LHC 
signatures of this scenario in great details.

In this paper we complement and extend the analysis in \cite{debottam} in several ways. First of all 
we have simulated the popular $m$-leptons + $n$-jets + $\etslash$ 
signature 
for $n$ $\ge$ 2 and several choices of $m$, which stands for the number of 
electrons and muons. No flavour tagging is required at this stage. It is 
gratifying to note that the scenarios can be distinguished reasonably well 
via these signatures inspite of the uncertainties in the cross-sections 
due to the choice of QCD scales.  The relative enhancement of the final 
states with $e$ and/ or $\mu$ compared to $m = 0$ final states, a distinct 
characteristic of the VLSP scenario, has been illustrated in 
the next section with several examples in scenario C.

Next we consider events of the type  $1\tau + X$,  $1\mu + X$ and $1 e + X$, 
where $X$ includes all possible states with two or more jets but no stable lepton or
tagged $\tau$ ( the difference in the definition of $X$ 
compared to that in \cite{debottam} 
should be noted). In this work we also consider final states with 2 
leptons of the same flavour in all possible charge combinations + $X$. We 
have used the $\tau$-tagging efficiencies provided by the CMS 
collaboration \cite{cms} in our Pythia based simulations. As discussed 
in the earlier paragraphs an excess of events with $\tau$ have been 
demonstrated in scenario A and to a lesser extent in scenario B. Finally we 
examine the $b$-jet content of the final states beyond the parton level using 
Pythia.

In \cite{debottam} the $t \bar{t}$ events were assumed to be the dominant 
source of backgrounds. In this paper we have extended the
background analysis by considering several other processes. In particular we have found that for some signals the QCD
background is more important than the $t \bar{t}$ background. The details 
will be presented in the next section.

\section{The Signals at the LHC}

\begin{table}[!htb]
\begin{center}\

\begin{tabular}{|c|c|c|c|}
       \hline
mSUGRA &A&B&C\\
parameters & &&\\
\hline
$m_0$ &120.0  &120.0 &120.0\\
\hline
$m_{1/2}$ &300.0  &350.0 &500.0\\
\hline
$A_0$ &-930.0  &-930.0 &0.0\\
\hline
$\tan\beta$ &10.0  &10.0 &10.0\\
\hline
$sign(\mu)$ &1.0  &1.0 &1.0\\
\hline

\end{tabular}
\end{center}
   \caption{Three bench mark scenarios introduced in \cite{debottam}.}
\end{table}

\begin{table}[!ht]
\begin{center}\
\begin{tabular}{|c|c|c|c|}
\hline
Decay modes & A &B&C\\
(Gauginos) & &&\\
\hline
$\lsptwo \ra \wt l_L^-  l^+ $    &  0.0 &  2.2 & 23.1\\
\hline
$\lsptwo \ra \wt l_R^-  l^+ $    &  0.9 &  0.6 &  0.8\\
\hline
$\lsptwo \ra \stau_1^{+} \tau^-$ & 78.6 & 46.0 &  8.4\\
\hline
$\lsptwo \ra \stau_2^{+} \tau^-$ &  0.0 &  0.7 & 10.2\\
\hline
$\lsptwo \ra \lspone  Z $        &  0.3 &  0.2 &  0.4\\
\hline
$\lsptwo \ra \lspone  h $        &  0.0 &  1.9 &  5.2\\
\hline
$\lsptwo \ra \snu_l \nu_l$       &  5.2 & 24.0 & 33.8\\
\hline
$\lsptwo \ra \snutau \nu_{\tau}$ & 15.0 & 24.0 & 17.8\\
\hline \hline
$\chonep \ra \lspone W^+$           & 2.6  &  2.6 &  5.1\\
\hline
$\chonep \ra \snu_l l^+$            & 5.4  & 25.0 & 36.0\\
\hline
$\chonep \ra \snutau \tau^+$        & 16.0 & 25.0 & 19.2\\
\hline
$\chonep \ra \wt l_L^+ \nu_l $      &  0.0 &  2.0 & 22.0\\
\hline
$\chonep \ra \stau_1^+ \nu_{\tau} $ & 76.0 & 44.0 &  7.8\\
\hline
$\chonep \ra \stau_2^+ \nu_{\tau} $ &  0.0 &  0.7 &  9.6\\
\hline
\end{tabular}
\end{center}
   \caption{The BRs of the dominant decay modes of the lighter chargino
and the second lightest neutralino. All sneutrinos decay into the invisible
channel $\nu$ + $\lspone$ in the three cases understudy. 
Here $l$ stands both for $e$ and $\mu$.}
\end{table}

\begin{table}[!ht]
\begin{center}\
\begin{tabular}{|c|c|c|c|c|c|c|c|}
\hline
Decay modes       & $S_{1}$ & $ S_{2} $ & $ S_{3} $& $ S_{4} $& $ S_{5} $& $ S_{6}$& $ S_{7}$ \\
(Gauginos) & &&&&&&\\
\hline \hline
$\lsptwo \ra \wt l_L^-  l^+ $    & 0.0    &  10.2 &  12.5 & 0.0    & 0.0    &  15.5  & 15.0\\
\hline
$\lsptwo \ra \wt l_R^-  l^+ $    & 1.6  &  0.7  &  0.5  & 0.9  & 0.7  & 0.2    & 0.2\\
\hline
$\lsptwo \ra \stau_1^{+} \tau^-$ & 85.8 &  26.6 &  24.2 & 95.4 & 76.9 & 19.7   & 20.2    \\
\hline
$\lsptwo \ra \stau_2^{+} \tau^-$ & 0.0    &  4.3  &  5.4  & 0.0    &   0.0  & 9.8    & 10.6     \\
\hline
$\lsptwo \ra \lspone  Z $        &  0.0 &  0.2  &  0.2  & 0.2  &  0.2  & 0.1    & 0.1\\
\hline
$\lsptwo \ra \lspone  h $        &  0.0    &  3.3  &  2.3 & 0.0 & 0.0    & 2.0    & 1.4\\
\hline
$\lsptwo \ra \snu_l \nu_l$       &  2.8    &  31.8 & 31.3  & 0.0 & 5.5  & 28.8   & 27.3 \\
\hline
$\lsptwo \ra \snutau \nu_{\tau}$ &  9.7    &  22.2 & 22.7  &3.5& 16.7 & 23.8   & 25.0 \\

\hline \hline
$\chonep \ra \lspone W^+$           & 1.3  &  3.6    &  2.5  & 1.9 & 2.2  &  2.0  &  1.5 \\
\hline
$\chonep \ra \snu_l l^+$            & 3.0  & 33.4    & 33.4  & 0.0 &5.7  & 29.6  &  28.0\\
\hline
$\chonep \ra \snutau \tau^+$        & 11.0 & 23.4    & 23.9  & 3.6 &17.7 & 24.5  &  25.7  \\
\hline
$\chonep \ra \wt l_L^+ \nu_l $      &  0.0 & 10.1    & 11.8  &  0.0  & 0.0 & 15.1  &  14.6 \\
\hline
$\chonep \ra \stau_1^+ \nu_{\tau} $ & 84.7 & 25.4    &  23.1 & 94.5 & 74.3  &19.2  &  19.8\\
\hline
$\chonep \ra \stau_2^+ \nu_{\tau} $ &  0.0 &  4.0    &  5.2  & 0.0    & 0.0 &9.5   &  10.3 \\
\hline
\end{tabular}
\end{center}
\caption{Same as Table 2 in the scenarios $ S_{1} - S_{7}$(see text).} 
\end{table}


In this section we begin with the generic SUSY signals of the type 
$ m-l+ n-j + \etslash $ ,where $l = e$ or $\mu $ and $j$ is any jet. At first
we do not employ any flavour tagging. Our aim is to study the 
feasibility of discriminating among the three models under consideration 
using these generic signals. Next we shall employ flavour tagging and 
demonstrate that it further enhances our discriminatory power.
        
The production and decay of all squark-gluino pairs are generated by 
Pythia (version 6.409) \cite{pythia}. Initial and final state radiation, 
decay, hadronization, fragmentation and jet formation are implemented 
following the standard procedures in Pythia. We have used the toy 
calorimeter simulation (PYCELL) provided in Pythia with the following 
criteria:

\begin{itemize}
\item The calorimeter coverage is $\vert \eta \vert < 4.5$. The segmentation is given by $\Delta \eta \times \Delta \phi = 0.09 \times 0.09$ which resembles a generic LHC detector.

\item A cone algorithm with $\Delta$ R$ = \sqrt {\Delta\eta^2 + \Delta\phi^2}= 0.5 $ has been used for jet finding.

\item E$^{\mathrm{jet}}_{\mathrm{T,min}} = 30 $GeV and jets are ordered in E$\mathrm{_T}$.
\end{itemize}

The stable leptons are selected according to the criterion :
\begin{itemize}

\item Leptons $(l=e,\mu)$ are selected with P$\mathrm{_T \ge 30}$ GeV
 and $\vert\eta \vert < 2.5$. For lepton-jet isolation 
we require $\Delta R(l,j) > 0.5$.The detection efficiency of the leptons are assumed to be $ 100 \%$.

\end{itemize}

The following cuts are implemented for background rejection :
\begin{itemize}

\item Leptons $(l=e,\mu)$ with P$\mathrm{_T \le 60}$ GeV are rejected to
ensure the rejection of leptons coming from $\tau$ decay.(CUT 1)

\item We reject events without at least two jets having P$\mathrm{_T} > 150$ 
GeV( CUT 2)

\item Events with missing energy ($\etslash) < 200$ GeV are rejected.(CUT 3)

\item Events with $M_{eff} < 1000$ are rejected,  
where $M_{eff}= |\met| + \Sigma_{i}|P_T^{l_i}| + \Sigma_{i}|P_T^{j_i}|$
($l = e,\mu$ ).(CUT 4)

\item Only events with jets having S$\mathrm{_T} > 2.0$, where 
S$\mathrm{_T}$ is a standard function of the eigenvalues of the 
transverse sphericity tensor,are accepted.(CUT 5)
\end{itemize}

\begin{table}[!ht]
\begin{center}\
\begin{tabular}{|c|c|c|c|}
\hline
QCD scale & $ 0.5 \sqrt{\hat s}$ & $ \sqrt{\hat s}$ & $ 2.0\sqrt{\hat s} $\\
\hline
A & 19.41 & 15.58 & 12.10 \\
\hline 
B &  6.79 &  5.74 &  4.26 \\
\hline
C &  0.84 &  0.74 &  0.51 \\
\hline
\end{tabular}
\end{center}
   \caption{Variation of total cross-section in $\pb$ of squrak-gluino events with the QCD scale in A, B and C.}

\end{table}
 
\begin{table}[!ht]
\begin{center}\
\begin{tabular}{|c|c|c|c|c|}
\hline
\multicolumn{2}{|c|}{Q}& $ 0.5 \sqrt{\hat s}$ & $ \sqrt{\hat s}$ & $ 2.0\sqrt{\hat s} $\\
\hline
A& $ 0 l$ & 2.54 &2.03  &1.59 \\
\hline
& $ 1 l$  & 0.25 &0.20  &0.15 \\
\hline
\hline
B& $ 0 l$ & 1.45 &1.21 &0.90 \\
\hline
& $ 1 l$  & 0.21 &0.17 &0.13 \\
\hline
\hline
C& $ 0 l$ & 0.24 &0.21 &0.15 \\
\hline
& $ 1 l$  & 0.11 &0.09 &0.06 \\
\hline

\end{tabular}
\end{center}
 \caption{The cross-sections (including efficiency) in $\pb$ of events with $m = 0$ and 
1 (see text).}
\end{table}

Cut 1 will be employed in the second part of our analysis when the relative 
abundance of final states with $e,\mu$ and $ \tau $ will be studied. For
establishing $ m-l+ n-j + \etslash $ signals Cut 2 - Cut 4 are adequate.

Table 4 illustrates that, as expected, the variation of the total 
cross-section of squark-gluino events with the choice of the QCD scale is 
rather large in each model. In spite of this large variation it is clear 
that each scenario is characterized by a typical size of the cross-section. 
This cross-section is the largest in scenario A because squark and 
gluino masses are the smallest in this case (see \cite{debottam}). In 
particular the contribution of the lighter top squark enhances the
cross-section significantly. The last feature is a direct consequence of 
large negative $A_0$ in scenario A. The corresponding masses are 
significantly larger in the other two scenarios resulting in smaller cross-sections.

It is, however, impossible to conclusively identify a particular 
DM scenario by the size of the cross-section alone since a similar 
cross-section may arise from a different combination of mSUGRA parameters 
which may or may not be allowed by the relic density constraint. Moreover, 
the signal corresponding to a larger raw cross-section may eventually be 
suppressed due to the effects of the kinematical cuts, small BRs of the 
underlying decays etc. Several examples of this will be presented in the 
following paragraphs.

The total cross-section  at best provides a hint for the underlying SUSY 
model and LSP annihilation mechanism but no definite conclusion can be drawn. 
In the following we shall show that a multi-channel analysis using signals with
 different choices of $m$ (the number of leptons in the final state) may very 
efficiently discriminate among different scenarios.

First we assume that the SM backgrounds will be determined either from 
data or from theory with next to leading order accuracy and can be 
subtracted from the event sample without introducing large errors.
The leading order signal cross-sections in $\pb$ (including the BRs and 
the efficiency of the cuts C2 -C5 listed above) for $m = 0$ and 1 are presented 
in Table 5. It is readily seen that the 
variation of the cross-sections due to the QCD scale, the difference in BRs 
(see section 2 and \cite{msugra} for further details) and the efficiency of 
the kinematical cuts in different scenarios may conspire in such a way 
that the cross-sections for $m = 0$ ($\sigma_0$) may look rather similar in 
different scenarios (see, e.g., $\sigma_0$ in model A at $Q = 2.0 
\sqrt{\hat s}$ and in model B at $Q = 0.5 \sqrt{\hat s}$)at least for 
relatively low integrated luminosities ($\cal L $). Similarly the predictions for 
$m = 1$($\sigma_1$) may appear  to be quite similar in different scenarios. 
However, it can be easily checked that in each 
scenario the ratio R=$\sigma_0 /\sigma_1$ is scale independent to a very 
good approximation. The value of R for $\cal L $ 
of 10 $\ifb$ in A, B and C are 10.3, 7.0 and 2.3 
respectively. Introducing a statistical uncertainty  $\sqrt N $ for
$ N $ counts and using the standard method for estimating the uncertainty
we find R$_{A} = 10.3 \pm 0.24$,  R$_{B} = 7.0 \pm 0.18$
and  R$_{C} = 2.3 \pm 0.9$. Thus it is fair to conclude that different scenarios
can be readily distinguished from each other.
Ratios involving cross-sections with other choices of $ m $ exhibit similar 
scale independence.  

The relatively large value of R in scenario A is partly due to the $\tau$ 
dominance of the final states in this scenario as discussed  in the last 
section. Since the $\tau$ decays into hadrons with a large BR, the 
numerator of R is naturally enhanced. The denominator of R on the other 
hand is small because, as explained in the last section, the number of 
final states involving $e$ and $\mu$ are suppressed.

The above properties also hold in scenario B 
to a lesser extent yielding a value of R smaller than that in A but 
significantly larger than the one in C. In C, as in any other VLSP 
scenario, the denominator of R is rather large for reasons already 
discussed and a smaller  R is obtained.

In order to provide some estimate of the dominant backgrounds we present in 
Table 6 the signal and the important standard model backgrounds 
for several values of $m$ in
the leading order for Q = $\sqrt {\hat s}$. This will be followed by the 
usual analysis 
of the significance ($ S/ \sqrt B$) , where $ S(B)$ is the total number of 
signal (background) events. Although we have listed only the
backgrounds from $ t \bar t$, QCD (including all quark-anti-quark and gluon 
events in Pythia in the lowest order ) we have also 
simulated $WW, ZZ, WZ $ and Drell-Yan backgrounds and have found them to be 
indeed negligible. For later use we have simulated the background from 
$W + jets$ events.

The $W + jets $ cross-section has been computed for $ \hat p_T > 50$ where
 $ \hat p_T$ is defined in the rest frame of the parton -parton collision.
The QCD  cross-section has been computed in two $ \hat p_T$ bins :
(i) 400 GeV $< \hat p_T <$ 1000 GeV   and 
(ii) 1000 GeV$< \hat p_T <$ 2000 GeV .The corresponding cross-sections
being 2090 $\pb$ and 10 $\pb$ respectively. Outside these bins
the number of events are negligible.  

Although the squark-gluino production cross-section is rather tiny in 
scenario C the signal cross-sections predicted for $m \geq 2$ is larger than 
the corresponding signals in A and B with much larger raw production cross-section. 
This again is a direct consequence of the large leptonic BRs of
gaugino decays in the underlying VLSP scenario as discussed above.
The significance (without systematic errors) of the signal 
for different $m$ and the representative $ \mathcal L $ of 10 $\ifb$ are 
in Table 7. The corresponding numbers for other $\cal L $ s 
can be estimated by simple scaling and one can easily verify that in 
several channels statistically significant signals can be obtained for 
much lower $\cal L$ s. 

From Tables 6 and 7 it is clear that one way to unambiguously discriminate
 between A, B on the one hand and C on the other 
is the count of $0-l$ events. This conclusion based on leading order cross-sections 
is likely to hold inspite of the scale uncertainty discussed above 
and the possibility that the systematic errors, which we have not considered 
in this paper, might be relatively large in the early stages of the LHC and 
affect the significance. The same count may 
discriminate between A and B although the theoretical uncertainties illustrated 
in Table 4 may cast some doubt on the results.
The observation of the clean almost background free signal for $m = 3$ may vindicate model C 
since no statistically significant signal is expected from A or B in this case. On the other
hand in the VLSP scenario C an acceptable signal in this channel is expected even for 
$ \mathcal L $ significantly smaller than 10 $\ifb$.

Counting experiments alone for $m = 1$ or $m = 2$ may not be very useful for discriminating 
between A and B during the early stages of the LHC run because
of the theoretical uncertainties and low statistics.
As discussed above one can form several QCD scale invariant ratios of observables to distinguish
between scenarios A and B. For example R$^\prime = \sigma_0 /\sigma_{2OS}$ ,where $\sigma_{2OS}$
is the cross-sections for $m = 2$ involving opposite sign leptons, R$^\prime_{A} = 312.3 \pm 38.7 $ and 
R$^\prime_{B} = 131.5 \pm 13.8 $. Another
example is  R$^{\prime \prime}= \sigma_1 /\sigma_{2SS}$ ,where $\sigma_{2SS}$
is the cross-sections for $m = 2$ same sign leptons, R$^{\prime \prime}_{A} = 83.3 \pm 17.1$ and
R$^{\prime \prime}_{B} = 50.0 \pm 8.7 $. However, more statistics will be required
to make the distinction unambiguous.

One could use the next to leading order cross-section 
for squark-gluino production \cite{zerwas}. However, the uncertainty in the dominant QCD background
in the canonical $0-l$ channel
due to higher order effects is not known precisely.
We have therefore restricted ourselves to leading order cross-sections 
for both the signal and the backgrounds. It is somewhat 
reassuring to note that the leading order signal cross-sections are typically multiplied by a factor of
1.4 - 1.5 in the next to leading order. Thus the significance will remain the same
even if we have underestimated the background by a factor of two.

Obviously the dominance of final states involving
$\tau$ leptons in some of the models under consideration can not be directly
established from the generic observables. Nor can the dominance of final states
with B-hadrons in certain scenarios be tested. For this $\tau$
and $b$-jet tagging facilities must be relied upon.

\begin{table}[!ht]
\begin{center}\
\begin{tabular}{|c|c|c|c|c|c|}
\hline
&\multicolumn{3}{c|}{SIGNAL} & \multicolumn{2}{c|}{background}\\
\hline
& A &B&C& $ t\bar t$ & QCD\\
\hline
$\sigma (\pb)$     & 15.58   & 5.74   & 0.74   & 400  & 2100\\
\hline \hline
$ 0 l $ & 2.03 & 1.21 & 0.21 & 0.33 & 3.55 \\
\hline
$ 1 l $ & 0.20 & 0.17 & 0.09 & 0.16 & $ 5.0 \times 10^{-3}$ \\ 
\hline
$ S S $ & $ 2.4 \times 10^{-3}$  & $ 3.4 \times 10^{-3}$ & $ 6.0 \times 10^{-3}$ & $7.2 \times 10^{-4}$& -\\
\hline
$ O S $ & $ 6.5 \times 10^{-3}$  & $ 9.2 \times 10^{-3}$ & 0.02 & 0.01 & -\\
\hline
$ 3l $ & $ 2.6 \times 10^{-4}$  & $ 4.9 \times 10^{-4}$ & $ 3.5 \times 10^{-3}$ &$ 3.2 \times 10^{-4}$& -\\
\hline
\end{tabular}
\end{center}
   \caption{The cross-sections (including efficiency) at $Q = \sqrt{\hat s}  $ for signal process
with different m, $t \bar t$ and QCD events. Here $SS$ refers to $m = 2$ with leptons carrying the 
same charge and $OS$ refers to similar events with leptons carrying opposite charge.
No entry in a particular column (-) means negligible background. }

\end{table}
\begin{table}
\begin{center}\
\begin{tabular}{|c|c|c|c|}
\hline
 & A & B & C \\
\hline
$ 0 l $ & 103 & 61  & 11  \\
\hline
$ 1 l $ & 49  & 42  & 22  \\
\hline
$  SS $ &  9  & 13  & 22  \\
\hline
$  OS $ &  5  &  8  & 16  \\
\hline
$ 3 l $ &  1  &  3  & 20  \\
\hline
\end{tabular}
\end{center}
   \caption{The significance (S/$\sqrt B$($t \bar t + QCD $))of signals in Table 6 for 
$ \mathcal L$$ = 10 \ifb $.}
\end{table}
\begin{table}[tbp]
\begin{center}\
\begin{tabular}{|c|c|c|c|c|c|c|}
\hline 
& \multicolumn{5}{c|}{Efficiencies for selection cuts} \\
\cline{2-6}
& Cut 1 &Cut 2&Cut 3& Cut 4  &Cut 5\\
\hline
$1 \tau + X$       & 0.0712  & 0.0441  & 0.0334   & 0.0263  & 0.0155\\
\hline
$1 e  + X$         & 0.0214  & 0.0084 & 0.0061  & 0.0051 & 0.0034\\
\hline
$1 \tau + 0 b + X$ & 0.0391  & 0.0273  & 0.0201   & 0.0161  & 0.0097\\
\hline
$1 \tau + 1 b + X$ & 0.0194  & 0.0112  & 0.0081   & 0.0062  & 0.0034 \\
\hline
$1 \tau + 2 b + X$ & 0.0113  & 0.0061  & 0.0041  & 0.0039   & 0.0022 \\
\hline
$1 e + 0 b + X$    & 0.0065 & 0.0030 & 0.0023  & 0.0019 & 0.0012\\
\hline
$1 e + 1 b + X$    & 0.0083 & 0.0031 & 0.0022  & 0.0019 & 0.0012\\
\hline
$1 e + 2 b + X$    & 0.0047 & 0.0019 & 0.0014  & 0.0012 & 0.0008\\
\hline 
                                                                                                                             
\end{tabular}
\end{center}
\caption{The cumulative efficiency of the cuts for signal(A) given by $N_i / N$, where $N_i$
is the number of  events survived after successive application of Cut 1 to Cut i and $N$ is the total
sample generated.}
                                                                                                                             
\end{table}
                                                                                                                             
                                                                                                                             
\begin{table}[tbp]
\begin{center}\
\begin{tabular}{|c|c|c|c|c|c|c|}
\hline 
& \multicolumn{5}{c|}{Efficiencies for selection cuts} \\
\cline{2-6}
& Cut 1 &Cut 2&Cut 3& Cut 4  &Cut 5\\
\hline
$1 \tau + X$        & 0.1374                  & 0.0135                & 0.0006               & 0.0006               &  1.29$\times 10^{-4}$\\
\hline
$1 e  + X$          &  3.7$\times 10^{-5}$  & 3.6$\times 10^{-5}$  & 2.7$\times 10^{-6}$ & 2.7$\times 10^{-6}$ &   1.9$\times 10^{-7}$\\
\hline
                                                                                                                             
$1 \tau + 0 b + X$  & 0.1261                  & 0.0125                & 2.7$\times 10^{-4}$  & 2.4$\times 10^{-4}$               &   4.6$\times 10^{-5}$\\
\hline
$1 \tau + 1 b + X$  & 0.0087                 & 0.0085                & 2.8$\times 10^{-4}$   & 2.5$\times 10^{-4}$               &  5.6$\times 10^{-5}$ \\
\hline
$1 \tau + 2 b + X$  & 0.0019                 & 0.0019                & 8.5$\times 10^{-5}$ & 7.9$\times 10^{-5}$ &  2.5$\times 10^{-5}$ \\
\hline
                                                                                                                             
$1 e + 0 b + X$     & 2.2$\times 10^{-5}$   & 2.1$\times 10^{-5}$  & 1.3$\times 10^{-6}$ & 1.3$\times 10^{-6}$ &  1.9$\times 10^{-7}$\\
\hline
$1 e + 1 b + X$     & 1.2$\times 10^{-5}$   & 1.2$\times 10^{-5}$  & 1.3$\times 10^{-6}$ & 1.3$\times 10^{-6}$ &  -\\
\hline
$1 e + 2 b + X$     & 2.0$\times 10^{-6}$   & 2.0$\times 10^{-6}$  &    -                &     -               &  -\\
\hline 
                                                                                                                             
\end{tabular}
\end{center}
\caption{Same as Table 8 for QCD events.}
\end{table}
                                                                                                                             
\begin{table}[tbp]
\begin{center}\
\begin{tabular}{|c|c|c|c|c|c|c|}
\hline 
& \multicolumn{5}{c|}{Efficiencies for selection cuts} \\
\cline{2-6}
& Cut 1 &Cut 2&Cut 3& Cut 4  &Cut 5\\
\hline
$1 \tau + X$        & 0.0392  & 0.0041  & 0.0006      & 0.0005              &  0.0001\\
\hline
$1 e  + X$          & 0.0481  & 0.0034  & 0.0004      & 0.0003              &  0.0001\\
\hline
$1 \tau + 0 b + X$  & 0.0101  & 0.0008  & 0.0002      & 0.0001              &  2.3$\times 10^{-5}$\\
\hline
$1 \tau + 1 b + X$  & 0.0192  & 0.0023   & 0.0003      & 0.0002              &  5.6$\times 10^{-5}$ \\
\hline
$1 \tau + 2 b + X$  & 0.0097 & 0.0013  & 0.0002      & 0.0001              &  3.0$\times 10^{-5}$ \\
\hline
$1 e + 0 b + X$     &  0.0133 & 0.0006  & 0.0001      & 6$\times 10^{-5}$  &  2.5$\times 10^{-5}$\\
\hline
$1 e + 1 b + X$     & 0.0244  & 0.0016  & 0.0002      & 0.0001              &  4.6$\times 10^{-5}$\\
\hline
$1 e + 2 b + X$     & 0.0113  & 0.0011  & 0.0001      &  9$\times 10^{-5}$ &  3.0$\times 10^{-5}$\\
\hline 
                                                                                                                             
\end{tabular}
\end{center}
\caption{Same as Table 8 for $t \bar t$ events.}
\end{table}

\begin{table}[!ht]
\begin{center}\
\begin{tabular}{|c|c|c|c|c|c|c|}
\hline
&\multicolumn{3}{c|}{SIGNAL} &\multicolumn{3}{c|}{BACKGROUNDS}\\
\cline{2-7}
& A &B&C& $ t\bar t$ &$ W + jets$& QCD  \\
\hline
$\sigma (\pb)$     & 15.58   & 5.74   & 0.74   & 400   & 650    & 2100\\
                                                                                                                             
\hline $1 \tau + X$& 0.2415   & 0.1251  & 0.0232 & 0.0433 & 0.0018  & 0.2709\\
\hline
$1 e  + X$         & 0.0530   & 0.0433  & 0.0302 & 0.0403 & 0.0015  & 0.0004\\
\hline
$1 \tau + 0 b + X$ & 0.1511   & 0.0744  & 0.0151 & 0.0091 & 0.0014  & 0.0958\\
\hline
$1 \tau + 1 b + X$ & 0.0530   & 0.0293  & 0.0041 & 0.0222 & 0.0003  & 0.1184\\
\hline
$1 \tau + 2 b + X$ & 0.0343   & 0.0212  & 0.0032 & 0.0120 & 0.0001  & 0.0524 \\
\hline
$1 e + 0 b + X$    &  0.0187  & 0.0141  & 0.0201 & 0.0101 & 0.0013   & 0.0004 \\
\hline
$1 e + 1 b + X$    & 0.0187   & 0.0152  & 0.0052 & 0.0182 & 0.0002  & - \\
\hline
$1 e + 2 b + X$    & 0.0125   & 0.0121  & 0.0043 & 0.0120 &  -       & -\\
\hline
\end{tabular}
\end{center}
   \caption{The cross-sections (including efficiency) of events with one detected $\tau$ 
and one isolated $e$. Here $X$ stands for all possible final states excluding any lepton 
or tagged $\tau$ but with at least two jets. 
The  number of tagged $b$-jets is given by $n$-$b$, $n=1,2,3$ (see the first 
column  of row 4 - 9)
.} 
\end{table}

We, therefore, turn our attention to final states of the type $1\tau + X$ 
where $X$ 
includes two or more hard jets but no $e$ or $\mu$ or tagged $\tau$. Tagging of 
$\tau$ jets are implemented according to the following procedure.  

Only hadronic $\tau$ decays 
are selected. The $\tau$-jets with $\eta <$3.0 are then divided into 
several $P_T$ bins. A $\tau$-jet  in any $P_T$ bin is then treated as tagged 
or 
untagged according to the efficiency ($ \epsilon_{\tau}$) given in
\cite{cms} figure 12.9 for a particular bin. In this analysis we have applied Cuts 1-5.
The corresponding efficiencies for various signals in scenario A and the
dominant QCD and $t \bar t$ backgrounds are listed in Tables 8-10.

The computation of  $1e + X$ type events are rather straight forward. 
 Here for simplicity we have assumed the 
e-detection efficiency to be 100 \%. In our generator level analysis the 
result for $1\mu + X$ is expected to be the same to a good approximation
and we do not present them separately. The 
number of events of the above two types subject to the kinematical cuts 
listed above are presented in the second and third rows of Table 11 along 
with the dominant SM backgrounds. The QCD background  
to  $1\tau + X$ events stems from mistagging of light flavour jets as $\tau$-jets. 
The mistagging probability has also been taken from \cite{cms} figure 12.9.

 The  $1\tau + X$  signal, if unambiguously observed,
 will disfavour model C. The size of the
 signal or the ratio N($1\tau + X$)/N($1 e + X$) can distinguish between
 scenarios A and B inspite of theoretical and possible
 systematic uncertainties. If this signal is not observed then the $1e + X$ signal can
 establish scenario C. If we require $X$ to be free of $b$-jets then the significance
 marginally increases because of the reduced background from $t \bar t$ events.
 The statistical significance of various
 signals for the representative $\cal L $ of 10 $\ifb$ are 
 listed in Table 12.

We next illustrate that both scenarios A and B are expected to be rich in
$b$-jets. A jet with $\vert \eta \vert < 2.5$ matching with a B-hadron of 
decay length $ > 0.9 \mathrm{mm} $ has 
been marked $ tagged$. The above criteria ensures that $\epsilon _{b} 
\simeq 0.5$ in $ t \bar t$ events, where $\epsilon _{b}$ is the 
single $b$-jet tagging efficiency (i.e., the ratio of the number of tagged 
$b$-jets and the number of taggable $b$-jets in $t \bar t$ events). 
It is readily seen that, the fraction of $1 e + X$ 
events with at least one tagged $b$-jet is quite large in A (0.59) and B (0.63) and 
somewhat smaller in C(0.31).
The significance of different signals are in Table 12. 

Finally we present in Table 13 events of the type $2\tau + X$ and $2 e + X$ 
with different number of tagged $b$-jets. It is to be noted that $2\tau + X$
events ($2 e + X$ events) are statistically significant in A and B (C).
Moreover an observable signal in 2$\tau$ +($\ge 1b)$ +$X$ channel is expected only in model A.

\begin{table}[!ht]
\begin{center}\
\begin{tabular}{|c|c|c|c|}
\hline
&\multicolumn{3}{c|}{SIGNAL}\\
\cline{2-4}
& A &B&C\\
\hline
$\sigma (\pb)$ & 15.58& 5.74 & 0.74\\
\hline
$1 \tau + X$       & 42.9  & 22.2  & 4.1\\
\hline
$1 e + X$          & 26.3  & 21.0  & 14.6\\
\hline
$1 \tau + 0 b + X$ & 46.4  & 22.7  & 4.6 \\
\hline
$1 \tau + (\ge 1 b) + X$ & 19.2  & 11.0  & 1.5 \\
\hline
$1 e + 0 b + X$ &  18.3   &  12.8 & 18.3\\
\hline
$1 e + ( \ge 1 b) + X$ & 17.9   & 15.6  & 5.4\\
\hline
\end{tabular}
\end{center}
   \caption{ The S/$\sqrt B$($t \bar t + W jets + QCD$) ratio for the signals in Table 11 corresponding to $ \mathcal L$$ = 10 \ifb $. }
\end{table}
\begin{table}[!ht]
\begin{center}\
\begin{tabular}{|c|c|c|c|c|c|}
\hline
&\multicolumn{3}{c|}{SIGNAL} &\multicolumn{2}{c|}{BACKGROUNDS}\\
\cline{2-6}
& A &B&C& $ t\bar t$ & QCD  \\
\hline
$\sigma (\pb)$     & 15.58    & 5.74    & 0.74   & 400      & 2100\\
\hline
$2 \tau + X$       & 0.0171     & 0.0078    & 0.0019   & 0.0015    & 0.0231\\
\hline
$2 e  + X$         & 0.0008    & 0.0009   & 0.0031   & 0.0010  & - \\
\hline
$2 \tau + 0 b + X$ & 0.0105    &0.0039    & 0.0012  &  0.0002  & 0.0084\\
\hline
$2 \tau + 1 b + X$ & 0.0038    &0.0019    & 0.0003  & 0.0007  & 0.0073\\
\hline
$2 \tau + 2 b + X$ & 0.0028    & 0.0017   & 0.0003  & 0.0006  & 0.0067\\
\hline
$2 e + 0 b + X$    & 0.0004    & 0.0004   & 0.0021  & 0.0002  & - \\
\hline
$2 e + 1 b + X$    & 0.0003    & 0.0003   & 0.0005  & 0.0006  & - \\
\hline
$2 e + 2 b + X$    & 0.0001    & 0.0002   & 0.0004  & 0.0002  & - \\
\hline
\end{tabular}
\end{center}
   \caption{The cross-sections (including efficiency)of events with two detected 
$\tau$ and two isolated $e$. The other conventions are as in Table 11.}
\end{table}

\begin{table}[!ht]
\begin{center}\
\begin{tabular}{|c|c|c|c|}
\hline
&\multicolumn{3}{c|}{SIGNAL}\\
\cline{2-4}
& A &B&C\\
\hline
$\sigma (\pb)$ & 15.58& 5.74 & 0.74\\
\hline
$2 \tau + X$       & 10.9  & 5.0  & 1.2 \\
\hline
$2 e + X$          & 2.5   & 2.8   & 9.8 \\
\hline
$2 \tau + 0 b + X$ & 11.5  & 4.3  & 1.3 \\
\hline
$2 \tau + (\ge 1 b) + X$ & 5.3   & 2.9  & 0.5 \\
\hline
$2 e + 0 b + X$    & 2.8   & 2.8   & 14.8 \\
\hline
$2 e + (\ge 1 b) + X$  & 1.4   & 1.8   & 3.2 \\
\hline
\end{tabular}
\end{center}
   \caption{The S/$\sqrt B$($t \bar t + QCD $) ratio of the signals in Table 13 for 
$ \mathcal L$ $ = 10 \ifb $. }
\end{table}

Although all the scenarios under consideration involve light sleptons 
direct slepton searches are unlikely to yield signals at the early LHC 
experiments. The slepton discovery plot in the $m_0$ - $\mhalf$ plane 
(see \cite{cms2} Fig.
13.29) shows that for $m_0$ = 120 GeV, our common choice for all three
scenarios, the reach in $\mhalf$ for 10 $\ifb$
is only about 150 GeV. However, for (50 - 60) $\ifb$ sleptons in scenarios A
and B should be detected. In the conventional $\stau$-annihilation scenario 
larger 
$ \mathcal L $ will be required for this discovery. The direct 
detection of sleptons will provide additional insight into the DM relic density
production.

Several earlier analyses of DM relic density in mSUGRA or other models considered
non-zero input values of $A_0$ \cite{severalnonzeroa0,stark}
with different emphasis 
Arnowitt et al in \cite{coannistau} showed that the $\stau$-coannihilation 
corridor in the $m_0$-$\mhalf$ plane is 
highly sensitive to $A_0$. For large 
$A_0$, $\stau$-coannihilation is effective even for $m_0$ as large as 1 TeV.
In a more recent work \cite{prannath} new bench mark points 
with non-zero $A_0$ allowed by WMAP data were introduced.
These points corresponds to new mass hierarchies and, 
consequently, new collider signatures. However, the correlation between 
leptonic signatures at LHC and different DM relic density
producing mechanisms was not discussed before.

In this paper we have considered the observed DM relic density constraints 
and direct constraints on sparticle masses from accelerators.
We have not considered indirect constraints like the  
measured value of BR( $B \ra  s \gamma$).
Many of the latter constraints  arise from flavour violating processes. 
Strictly speaking these constraints are 
sensitive to the assumption that the quark and the squark mass matrices are 
aligned in the flavour space so that the same 
mixing matrix as the CKM matrix operate in the squark sector. This assumption of minimum flavour violation
fails even if there are small off-diagonal elements of the
squark mass matrix at the GUT scale. On the other hand such small elements does not
affect processes like neutralino annihilation and squark-gluino 
production and decay. For further
discussions on this point we refer the reader to Djouadi et al in \cite{dmsugra}
and references there in. We, note in passing that model B and model C are allowed
by the above constraint (see \cite{debottam}).

\section{Summary and Conclusions}

~~~We have examined the parameter space of the mSUGRA model with moderate 
values of the 
parameter tan $\beta$. We focussed on zones of the $m_0 -\mhalf$ plane 
corresponding to 
light sleptons and relatively light squark and gluinos compatible with the 
DM relic 
density data and constraints from direct sparticle searches. This part of the 
parameter space is interesting since viable signals from squark-gluino events 
are expected in the early stages of the LHC experiments.

If one employs the often used but rather ad hoc assumption that the common 
trilinear 
coupling ($A_0$) vanishes at the GUT scale it is well known that there is only 
one such 
zone where $\stau$-coannihilation is the dominant mechanism for producing 
the relic 
density \cite{dmsugra}. In this scenario the $\chonepm$ and $\lsptwo$ present 
in squark 
gluino decay cascades exclusively decay via two body modes into appropriate 
neutrino (sneutrino) - slepton (lepton) pairs and
lepton-slepton, neutrino-sneutrino pairs. 
Lepton flavour universality holds in these decays to a very good approximation. The 
sneutrinos decay into invisible channels with almost 100\% BR. The $\lsptwo$ also decays 
into invisible channels with large BRs. The suppression of hadronic decays of the lighter 
electroweak gauginos and the presence of additional carriers of missing energy lead to 
spectacular collider signals as has already been noted in the context of LEP/NLC 
\cite{addrees,adaksr} and Tevatron \cite{ admgbm}. In this paper we emphasize that this 
VLSP 
or ELSP scenario is realized in the popular $\stau$-coannihilation region of the mSUGRA 
model and discuss the signatures at the LHC in detail by introducing the benchmark scenario 
C (see Table 1).

If the ad hoc assumption of vanishing $A_0$ is given up additional WMAP allowed parameter 
spaces open up \cite{debottam}. It is possible that the LSP pair annihilation and $\stau$- 
coannihilation both contribute significantly to DM relic density production although the 
former dominates. To illustrate the collider signals in this case the bench mark scenario A 
(see Table 1) is introduced. It is also possible that even with non-zero $A_0$ the $\stau$ 
coannihilation is the dominant mechanism but the corresponding squark-gluino masses are 
much smaller compared to C (see scenario B, Table 1). In scenario A and, to a lesser 
extent, in scenario B both $\chonepm$ and $\lsptwo$ decay dominantly into final states 
involving $\tau$ leptons (see the discussions in section 2 and Table 2) and lepton universality is violated. 
Moreover, these final states from squark-gluino production should be rich in $b$-jets.

In \cite{debottam} the collider signals from squark-gluino events in three scenarios were 
compared and contrasted by exploiting the above characteristics. The calculations were 
done mostly at the parton level although in order to simulate the effects of $\tau$ tagging 
some Pythia based analyses were also reported.

In this paper we first employ the generic $m$-leptons + $n$-jets + $\etslash$ signatures to
discriminate among the three scenarios ( see Tables 6 and 7), where only 
final states with stable leptons have been considered. We demonstrate that
some qualitative idea about the DM relic density producing mechanisms may be obtained even 
without flavour tagging. It is shown that the fraction of events with $m = 0$ is much larger
than that for $m \geq 1$ in scenarios A and B. The relative weight of the leptonic events,
 however, is significantly larger in scenario C (see Tables 5 - 7).

Next we illustrate the $\tau$ dominance of the final states in A and B by including $\tau$ 
detection efficiency in our simulation (see Tables 11 -14). Here we have extended the 
analysis of \cite{debottam} by considering $2l +X$ states where $l$ stands for e and 
tagged $\tau$ and $X$ corresponds to hadronic states. In particular the observation of 
$2\tau + X$ events may provide a very convincing test of scenario A. The number of tagged 
$b$-jets may further help to discriminate among different scenarios.
Most of the crucial 
signatures discussed in this paper may be observed with $ \mathcal L$ of 10 
 $\ifb$. Some of them are observable with much smaller accumulated luminosity.\\

{\bf Acknowledgment}: 
AD, SP and NB acknowledge financial support from  Department of Science and 
Technology, Government of India under the project  No (SR/S2/HEP-18/2003).
SP also thanks the Council of Scientific and Industrial Research (CSIR), 
India for a research fellowship. A large part of this work was done when the 
authors were in the Department of Physics, Jadavpur Universiry, 
Kolkata 700 032, India.

\end{document}